\documentclass[twoside]{article}
\usepackage[latin1]{inputenc}
\usepackage{amssymb}

\usepackage[german,english]{babel}
\usepackage{multicol}
\usepackage{vmargin}
\setpapersize{A4}
\setmargnohf{35mm}{45mm}{132mm}{220mm}

\begin{document}

\title{A Machine-Independent Port of the MPD Language Run Time System to
NetBSD Operating System} 

\author{Ignatios Souvatzis\\
	University of Bonn, Computer Science Department, Chair V\\ {\tt
	<ignatios@cs.uni-bonn.de>}}
\maketitle
\thispagestyle{empty}

\section{Introduction}

MPD (presented in Gregory
Andrews' book about Foundations of Multithreaded, Parallel, and
Distributed Programming\cite{MPDbook}) is the successor of 
SR\cite{srbook} (``synchronizing resources''), a PASCAL--style
language enhanced with constructs for concurrent programming developed
at the University of Arizona in the late 1980s\cite{srold}.

MPD as implemented provides the same language primitives as SR with a
different syntax which is closer to C.

The run-time system (in theory, identical) of both languages provides
the illusion of a multiprocessor machine on a single single-- or
multi--CPU Unix--like system or a (local area) network of Unix-like machines. 

\begin{table}
\begin{center}
\begin{tabular}{|c|c|c|}
\hline test machine & A & B \\
\hline
architecture & i386 & arm \\
CPU & Pentium 4 & SA-110 \\
clock & 1600 MHz & 233 MHz \\
cache & 2 MB & 16kB I + 16 kB D \\
\hline
\end{tabular}
\end{center}
\caption{\em Test machines}
\label{testmachine}
\end{table}

Chair V of the Computer Science Department of the University of Bonn
is operating a laboratory for a practical course in parallel programming
consisting of computing nodes running NetBSD/arm, normally used via PVM,
MPI, etc.

We are considering to offer SR and MPD for this, too. As the original
language distributions are only targeted at a few commercial Unix
systems, some porting effort is needed, outlined in the SR porting
guide\cite{sr:portingguide} and also applicable to MPD.

The integrated POSIX threads support of NetBSD-2.0 enables us to use
library primitives provided for NetBSD's pthread system to implement 
the primitives needed by the SR and MPD run-time systems, thus
implementing 13 target CPUs with a one-time effort; once implemented,
symmetric multiprocessing (SMP) would automatically be used on any 
multiprocessor machine with VAX, Alpha, PowerPC, Sparc, 32-bit Intel
and 64 bit AMD CPUs.

This paper describes mainly the MPD port. Porting SR was started
earlier and partially described in \cite{souvatzis:sr:port}
(Assembler and SVR4 cases) while
only preliminary results for our new approach could be presented at the
conference.

Most of the differences between our changes to SR and to MPD could be
done by mechanically replacing
{\tt mpd\_} by {\tt sr\_} in the code; because of this, and because
the machine-independent parts of the SR and MPD run-time support are
identical (according to the authors) all results (especially timing
results) equally apply to the SR port. (This has been verified.)

\section{Generic Porting Problems}

Despite the age of SR, the latest version (2.3.3) had been changed to
use {\tt <stdarg.h>} instead of {\tt <varargs.h>},
thus cutting the number of patches needed for
NetBSD 2.0 and later by half compared to the original porting effort
described in \cite{souvatzis:sr:port}. MPD 1.0.1 contains no traces
of {\tt <varargs.h>}. 


The only patches -- outside of implementing the context switching routines --
were for 64 bit cleanliness (see also \cite{husemann:lemmings}).

\section{Verification methods}

MPD itself provides a verification suite for the whole system;
also a small basic test for the context switching primitives.
There is no split between the basic and the extended verification suite,
as in SR.

\subsection{Context Switch Primitives}

The context switch primitives can be independently tested by running
{\tt make} in the subdirectory {\tt csw/} of the distribution; this
builds and runs the {\tt cstest} program, which implements a small
multithreaded program and checks for detection of stack overflows, 
stack underflows, correct context switching etc.\cite{sr:portingguide}
This test is automatically run when building the whole system.

\subsection{Overall System}

When the context switch primitives seem to work individually, they need 
to be tested integrated into the run-time system. The SR and MPD authors
provide a verification suite in the {\tt vsuite/} subdirectory of the
distributions to achieve this, as well as testing the the building system
used to build MPD, and the {\tt mpd} compiler, {\tt mpdl} linker, etc.

It is run by calling the driver script
{\tt mpdv/mpdv}, which provides options for selecting normal vs.\ verbose
output, as well as selecting the installed vs.\ the freshly compiled
MPD system. 

\begin{table}
\begin{center}
\begin{tabular}{|c|c|c|}
\hline Implementation & A & B \\
\hline
assembler & 0.013 $\mu$s& n/a\\
\ldots{}context\_u libary calls & 0.138 $\mu$s& 0.237 $\mu$s\\
SVR4 system calls & 1.453 $\mu$s&  9.649 $\mu$s\\
\hline
\end{tabular}
\end{center}
\caption{\em Raw context switch times}
\label{cswtimes}
\end{table}

For all porting methods described below (assembler primitives, SVR4 
system calls and NetBSD pthread library calls), the full verification
suite has been run and any reported problem has been fixed. 

\section{Performance evaluation}

MPD comes with two performance evaluation packages. The first, for the
context switching primitives, is in the {\tt csw/} subdirectory of 
the source distribution; after {\tt make csloop} you can start
{\tt ./csloop N} where N is the number of seconds the test will run
approximately.

Tests of the language primitives used for multithreading are in the
{\tt vsuite/timings/} subdirectory of the source tree enhanced with 
the verification suite. They are run by three shell scripts used to 
compile them, executed them, and summarize the results in a table.

\section{Establishing a baseline}

There are two extremes possible when implementing the context switch
primitives needed for MPD: implementing each CPU manually in assembler code
(what the MPD implementation does normally) and using the SVR4-style
functions {\tt getcontext()}, {\tt setcontext()} and {\tt swapcontext()}
which operate 
on {\tt struct ucontext}; these are provided as experimental code
in the file {\tt csw/svr4.c} of the MPD distribution.

The first tests were done by using the provided i386 assembler
context switch routines. After verifying correctness and noting 
the times (see tables \ref{cswtimes} and \ref{hltimes}), the same
was done using the SVR4 module instead of the assembler module.

These tests were done on a Pentium 4 machine running at 1600 MHz
with
2 megabytes  of secondary cache,
and 1 GB of main memory, 
running NetBSD-3.0\_BETA as of end of October 2005.

\begin{table}
\begin{center}

\begin{tabular}{|c|r|r|r|}
\hline Test description & i386 ASM & \ldots{}context\_u & SVR4 s.c.\\
\hline
loop control overhead			& 0.002 $\mu$s& 0.002 $\mu$s& 0.002 $\mu$s\\
local call, optimised			& 0.011 $\mu$s& 0.011 $\mu$s& 0.011 $\mu$s\\
interresource call, no new process 	& 0.270 $\mu$s& 0.260 $\mu$s& 0.250 $\mu$s\\
interresource call, new process 	& 0.650 $\mu$s& 4.200 $\mu$s& 4.350 $\mu$s\\
process create/destroy 			& 0.540 $\mu$s& 4.020 $\mu$s& 4.280 $\mu$s\\
\hline
semaphore P only 			& 0.011 $\mu$s& 0.011 $\mu$s& 0.011 $\mu$s\\
semaphore V only 			& 0.008 $\mu$s& 0.008 $\mu$s& 0.008 $\mu$s\\
semaphore pair 				& 0.019 $\mu$s& 0.019 $\mu$s& 0.019 $\mu$s\\
semaphore requiring context switch 	& 0.110 $\mu$s& 0.220 $\mu$s& 1.550 $\mu$s\\
\hline
asynchronous send/receive 		& 0.300 $\mu$s& 0.290 $\mu$s& 0.300 $\mu$s\\
message passing requiring context switch & 0.400 $\mu$s& 0.560 $\mu$s& 1.920 $\mu$s\\
\hline
rendezvous 				& 0.600 $\mu$s& 0.850 $\mu$s& 4.200 $\mu$s\\
\hline
\end{tabular}
\end{center}

\caption{\em Run time system performance, system A (Pentium 4, 1600 MHz).
The median times reported by
the MPD script {\tt vsuite/timings/report.sh} are shown.}
\label{hltimes}
\end{table}

The SVR4 tests were redone on a DNARD system (for its ARM cpu, no
assembler stubs are provided in either the SR or MPD distributions).

Table \ref{hltimes} shows a factor-of-about-ten performance hit for
the operations
that require context switches; note, however, that the absolute values
for all such operations are still smaller than $5\,\mu{}s$ on 1600\,MHz 
machine and will likely not be noticeable if a parallelized program is 
run on a LAN-coupled cluster: on the switched LAN connected to the test
machine, the time for an ICMP echo request to return is about 200 $\mu{}s$.

\begin{table}
\begin{center}
\label{hltimesarm}

\begin{tabular}{|c|c|r|r|}
\hline Test description & ARM ASM & \ldots{}context\_u & SVR4 s.c.\\
\hline
loop control overhead		& n/a & 0.057 $\mu$s& 0.056 $\mu$s\\
local call, optimised		& n/a & 0.376 $\mu$s& 0.355 $\mu$s\\
interresource call, no new process & n/a & 4.300 $\mu$s& 4.080 $\mu$s\\
interresource call, new process  & n/a & 27.250 $\mu$s& 55.900 $\mu$s\\
process create/destroy 		& n/a & 25.240 $\mu$s& 58.780 $\mu$s\\
\hline
semaphore P only 		& n/a & 0.304 $\mu$s& 0.301 $\mu$s\\
semaphore V only 		& n/a & 0.254 $\mu$s& 0.249 $\mu$s\\
semaphore pair 			& n/a & 0.506 $\mu$s& 0.487 $\mu$s\\
semaphore requiring context switch & n/a & 1.570 $\mu$s& 11.180 $\mu$s\\
\hline
asynchronous send/receive 	& n/a & 5.550 $\mu$s& 5.190 $\mu$s\\
message passing requiring context switch & n/a & 6.740 $\mu$s& 30.140 $\mu$s\\
\hline
rendezvous 			& n/a & 9.600 $\mu$s& 54.000 $\mu$s\\
\hline
\end{tabular}
\end{center}

\caption{\em Run time system performance, system B (StrongARM SA-110, 233 MHz).
The median times reported by the MPD script {\tt vsuite/timings/report.sh} are shown.
}
\end{table}

\section{Improvements using NetBSD library calls}

While using the system calls {\tt getcontext} and {\tt setcontext}, as
the {\tt svr4} module does, should not unduly penalize an application
distributed across a LAN, it might be noticeable with local applications.

However, we should be able to do better than the {\tt svr4} module without
writing our own assembler modules, since NetBSD 2.0 (and later) contains its
own set of them for the benefit of its native Posix threads library 
({\tt libpthread}), which does lots of context switches within a kernel
provided light weight process\cite{nathan:sa}. The primitives
provided to {\tt libpthread} by its machine dependent part are the three
functions {\tt \_getcontext\_u}, {\tt \_setcontext\_u} and
{\tt \_swapcontext\_u} with similar signatures as the SVR4-style 
system calls {\tt getcontext},
{\tt setcontext} and {\tt swapcontext}.

There were a few difficulties that arose while pursuing this.

First, on one architecture (i386) {\tt \_setcontext\_u} and {\tt
\_getcontext\_u} are implemented by calling through a function pointer
which is initialized depending on the FPU / CPU extension mode available
on the particular CPU used (8087-mode vs. XMM). On this architecture,
{\tt \_setcontext\_u} and {\tt \_getcontext\_u} are
defined as macros in a private header file not installed. The developer
in charge of the code has indicated that he might implement public
wrappers; until then, we have to check all available NetBSD
architectures and copy the relevant code to our module {\tt csw/netbsd.c}.

Second, we need to extract the relevant object
modules from the threading library for static linking ({\tt libpthread.a})
without resolving any other symbols,
because normal libpthread is overloading some system calls thus causing
failure of applications not properly initializing it.

Again, this set of context switch code has been verified by running
{\tt cstest} and the full verification suite.

The low-level as well as the high-level timings with the new context
switch package have again been collected in tables \ref{cswtimes}, 
\ref{hltimes} and \ref{hltimesarm}.

To ease installation, a package for the NetBSD package system
has been built for SR and MPD, available in the {\tt lang/sr} and
{\tt lang/mpd} subdirectories of the pkgsrc root.

As the NetBSD package system is available for more operating systems
than Net\-BSD\cite{crooks:portablepkg},
a lot more work would be needed to make the packages
universal; thus they are restricted to be built on NetBSD 2.0 and later.

\section{Discussion}

Our new approach has raw context switch times
that are only 10\% of the SVR4 system call ones. Compared to the 
assembler routines, they are only slower by a factor of 10 (see 
table \ref{cswtimes}).

Table \ref{hltimes} shows three classes of high level operations.

\begin{enumerate}

\item Non-context switching operations have the same speed independent
of the context switch primitives used, as expected. 

\item The two operations measured requiring a process creation (in
the MPD language sense) are about as fast as in the SVR4-system-call
case. This was expected, as the process creation primitive does a 
system call internally.

\item Context switching operations which do not create a new process
(in the MPD language sense) are slower than in the assembler case, but
faster than in the SVR4-style case, by an amount roughly equivalent
to one (semaphore operation, message passing) or two (rendezvous)
context switching primitive times.

\end{enumerate}

The same classification can be done for the 233 MHz ARM CPU (table
\ref{hltimesarm}). However, SVR4 process creation, destruction
and the rendezvous need about one third of the LAN two-way network 
latency, thus cannot
be neglected anymore. We conclude that for machines in the 300 MHz range
and below, using assembler implementation (where available) or at least
our new implementation of the context switching primitives is a necessity.
This is also expected for even slower machines.

MPD can be compiled in a mode where it will make use of multiple
threads provided by the underlying OS, so that it can use more than one
CPU of a single machine. This has not been implemented yet for NetBSD,
but should be.

\section{Summary}

A method for porting SR and MPD to NetBSD has been shown, for which
only preliminary results, and only for SR, were presented earlier.

The SR porting effort was easily adopted for the MPD case. In fact,
the run time system (library and srx/mprx) could probably be
factored out into a common run-time system package.

The new port was verified using the SR and MPD verification suites.

As discussed above, the SVR4-system-call approach, while feasible,
creates an overhead that is clearly visible for non-networked operation
of a distributed program; on our Pentium machine, high level context
switching operations are slower by a factor between 7 and 11 (the raw
context switch primitives are slower by a factor of 110). Even for 
networked operation, for a 233 MHz StrongArm CPU or slower machines, 
context switch latency exceeds one third of the network latency.

The approach using the libpthread primitives is much faster for all but
the process creation/destruction case and should thus be adequate
for about any application in the networked case, and for any in the
single-machine case that does not do excessive amounts of implicit or
explicit process creation.

For highly communication-bound problems on a single machine, using
the assembler primitives might show a visible speedup, where available.

\end{document}